\documentstyle[aps,prl,twocolumn,graphics,epsfig,floats]{revtex}

\newcommand{\be}{\begin{eqnarray}}
\newcommand{\ee}{\end{eqnarray}}

\begin{document}

\title{
The no-signaling condition and quantum dynamics }

\author{Christoph Simon$^{1}$\thanks{formerly: Institut f\"ur
Experimentalphysik, Universit\"at Wien, Boltzmanngasse 5, A-1090
Wien,  Austria}, Vladim\'{\i}r Bu\v{z}ek$^{2,3}$,
 and Nicolas Gisin$^4$ }
\address {$^1$
Centre for Quantum Computation, University of Oxford, Parks Road,
Oxford OX1 3PU, United Kingdom\\  $^2$ Institute of Physics, Slovak
Academy of Sciences, D\'{u}bravsk\'{a} cesta 9, 842 28 Bratislava,
Slovakia\\ $^3$ Faculty of Informatics, Masaryk University,
Botanick\'{a} 68a, 602 00 Brno, Czech Republic\\$^4$Groupe de
Physique Appliqu\'{e}e, Universit\'{e} de Gen\`{e}ve, 20 rue de
l'Ecole de M\'{e}decine, 1211 Gen\`{e}ve 4, Switzerland\\ }

\date{\today}
\maketitle

\begin{abstract}
We show that the basic dynamical rules of quantum physics can be
derived from its static properties and the condition that
superluminal communication is forbidden. More precisely, the fact
that the dynamics has to be described by linear completely positive
maps on density matrices is derived from the following assumptions:
(1) physical states are described by rays in a Hilbert space, (2)
probabilities for measurement outcomes at any given time are
calculated according to the usual trace rule, (3) superluminal
communication is excluded. This result also constrains possible
non-linear modifications of quantum physics.

\end{abstract}

The special theory of relativity is one of the cornerstones of our
present scientific world-view. One of its most important features
is the fact that there is a maximum velocity for signals, i.e. for
anything that carries information, identical to the velocity of
light in vacuum.

Another cornerstone of our present understanding of the world is
quantum physics. Quantum physics seems to have ``nonlocal''
characteristics due to the existence of entanglement. Most
importantly, it is not compatible with local hidden variables, as
shown by the violation of Bell's inequalities \cite{bell}, which
has been experimentally confirmed in many experiments
\cite{aspect,tittel,weihs}.

It is very remarkable that in spite of its non-local features,
quantum mechanics is compatible with the special theory of
relativity, if it is assumed that operators referring to space-like
separated regions commute. In particular, one cannot exploit
quantum-mechanical entanglement between two space-like separated
parties for communication of classical messages faster than light.

On the other hand, if one tries to modify quantum physics, e.g. by
introducing non-linear evolution laws for pure states
\cite{weinberg}, this easily leads to the possibility of
superluminal communication
\cite{gisinweinberg,gisinhpa,czachorweinberg}.

This peaceful, but fragile, coexistence between quantum physics and
special relativity has led physicists to ask whether the principle
of the impossibility of superluminal communication, which we will
refer to as the ``no-signaling condition'', could be used as an
axiom in deriving basic features of quantum mechanics. The answer
to this question should at the same time provide insight into what
kind of modifications of quantum physics are compatible with the
no-signaling condition.

Here we give a positive answer to the above question. If the usual
{\it static} characteristics of quantum mechanics are assumed, then
its {\it dynamical} rules can be derived from the no-signaling
assumption. By static characteristics we mean the following. The
states of our systems are described by vectors in a Hilbert space.
Furthermore at any given time we have the usual observables
described by projections in the Hilbert space \cite{nlobs}, and the
probabilities for measurement results are calculated according to
the usual trace rule \cite{peres}. However, no a priori assumption
is made about the time evolution of the system. For example, the
states could evolve according to some non-linear wave equation.
Note that we also do not assume the projection postulate.

 Our result is then that under the stated conditions the dynamics of our
system has to be described by {\it completely positive (CP) linear}
\cite{preskill} maps on {\it density matrices}. This is equivalent
to saying that under the given assumptions quantum physics is
essentially the only option since according to the Kraus
representation theorem \cite{preskill}, every CP map can be
realized by a quantum-mechanical process, i.e. by a linear and
unitary evolution on a larger Hilbert space. On the other hand, any
quantum process corresponds to a CP map. This result is a
significant extension of earlier work by one of the authors
\cite{gisinhpa}.

Let us first recall how the linearity of standard quantum dynamics
prevents the use of entanglement for superluminal communication.
Consider two parties, denoted by Alice and Bob, who are space-like
separated, such that all operations performed by Alice commute with
all operations performed by Bob. Throughout this work we will
assume that in {\it this} sense locality is implemented in the
algebra of observables. For example, this is certainly the case in
relativistic quantum field theory. Can the two parties use a shared
entangled state $\psi_{AB}$ in order to communicate in spite of
their space-like separation?

The short answer is: no, because the situation on Alice's side will
always be described by the same reduced density matrix,
irrespective of Bob's actions. All the effects of his operations
(described by linear maps) disappear when his system is traced
over. This answer is correct, but not very detailed, and thus it
may not be entirely convincing. Let us give a more detailed answer
which highlights the essential role played by the linearity of
quantum dynamics.

A question that is frequently raised in this context is the
following: Bob could choose to measure his system in two different
bases and thus ``project'' Alice's system into different pure
states depending on the basis he chose and his measurement result.
Since it is possible to distinguish two different states in quantum
mechanics, at least with some probability, shouldn't it be possible
for Alice to infer his choice of basis, at least in some percentage
of the cases, which would be dramatic enough?

Of course, the answer is no again, for the following reason. In
order to gain information about which basis Bob chose to measure,
Alice can only perform some (generalized) measurement on her
system. Then she has to compare the conditional probabilities for a
given result to occur, for the case that Bob measured in the first
or in the second basis. But these conditional probabilities will
always be exactly the same for both possibilities.

This can be seen as a consequence of the linearity of the quantum
physical time evolution: Suppose that Bob's first choice projects
Alice's system into states $\psi_i$ with probabilities $p_i$ and
his second choice projects it into states $\phi_k$ with
probabilities $q_k$. Alice can calculate the probability for her
obtained result for every one of the states, and then weight these
probabilities with the probability to have this specific state. But
because of the linearity of any operation that Alice can perform on
her states during her generalized measurement procedure, her final
result will only depend on the density matrix of the probabilistic
mixtures, which is the same in both cases, because they were
generated from the same entangled state. For an example how two
such mixtures can become distinguishable through a non-linear
(non-quantum-mechanical) evolution, see \cite{gisinweinberg}.

Let us note that this argument also implies the non-existence of a
perfect cloner in quantum mechanics because such a machine would
allow superluminal communication \cite{cloning} by making it
possible for Alice to discriminate between Bob's choices of basis.

We now show how quantum dynamics can be derived from ``quantum
statics'' and the no-signaling condition. As explained above, by
quantum statics we mean that physical systems are described by rays
in a Hilbert space, that at any given time we have the usual
quantum observables represented by projections in this Hilbert
space, and that the trace rule for calculating probabilities holds.

From the trace rule it follows that at any given moment the results
of measurements on some system $A$ are determined by its density
matrix or reduced density matrix, depending on whether $A$ is in a
probabilistic mixture of pure states or entangled with some other
system.

A priori we make no assumption about the dynamics of pure states,
e.g. it might be described by a non-linear wave equation. Let us
note immediately that, if the pure states have a non-linear time
evolution, then the density matrix of a probabilistic mixture is
not sufficient to determine the dynamics of the system, one has to
know the individual pure states and their probabilities.

If we consider a subsystem of the whole Universe it will in general
be in an entangled state with other parts of the Universe. In
particular, a system $A$ may be entangled with another system $B$
which is {\em space-like} separated from $A$, such that their
observable algebras commute. This is where the no-signaling
constraint comes into play. The dynamics of the systems has to be
such that in spite of this entanglement no superluminal
communication between $A$ and $B$ is possible.

Suppose that $A$ and $B$ together are in the entangled state
$|\psi\rangle_{AB}$ with reduced density matrix $\rho_A$ for system
$A$. As a consequence of the trace rule, by performing a
measurement of his system  the observer $B$ also prepares a certain
state in $A$.

To see this, remember that the trace rule tells us how to calculate
the (joint) probability for measurement results corresponding to
any product of projectors $P_A
\otimes P_B$, namely by calculating $\mbox{Tr}_{AB} \rho_{AB} P_A
\otimes P_B$. But this also tells us how to calculate the
conditional probability to find any $P_A$, provided that $P_B$ has
been found. Namely, we just have to divide the joint probability by
the probability to find $P_B$ in the first place. But having a way
of calculating the conditional probability for every $P_A$ means
that we know the state in $A$ conditional on $B$ having found
$P_B$, since a state can be reconstructed from its expectation
values for a linearly independent set of projectors. It is given by
$\mbox{Tr}_B
\rho_{AB} P_B / \mbox{Tr}_{AB}
\rho_{AB} P_B$. Note that to arrive at this conclusion we did not have to make use
of the usual projection postulate.

Actually, {\em every} probabilistic mixture of pure states
corresponding to the density matrix $\rho_A$ can be prepared via
appropriate measurements on $B$ \cite{gisinhpa,hughston}. We will
give a proof of this statement in the last part of this letter.

Consider two such probabilistic mixtures $\{P_{\psi_i}, p_i\}$ and
$\{P_{\phi_j}, q_j\}$, where $P_{\psi_k}$ is the projector
corresponding to the pure state $|\psi_k\rangle$ and $p_k$ is its
probability, such that
\begin{equation}
\sum \limits_i p_i P_{\psi_i}=\sum \limits_j q_j
P_{\phi_j}=\rho_A. \label{xx}
\end{equation}

According to the no-signaling principle there should be no way for
the  observer in $A$ to distinguish these different probabilistic
mixtures.

A general dynamical evolution in system $A$ is of the form
\be
g: P_{\psi}\rightarrow g(P_{\psi}) \ee where, most importantly, $g$
is not necessarily linear. Furthermore, $g(P_{\psi})$ does not have
to be a pure state, since system $A$ could become entangled with
its environment, or $\psi$ could evolve into a probabilistic
mixture of pure states. As mentioned above, even if system $A$ is
entangled with its environment, the trace rule implies that at any
given moment the results of measurements on $A$ will be completely
determined by the reduced density matrix of the system. In this
case we define $g(P_{\psi})$ to be the reduced density matrix of
$A$. If $\psi$ evolves into a probabilistic mixture, we define
$g(P_{\psi})$ to denote the corresponding density matrix.

Under such dynamics the probabilistic mixture $\{P_{\psi_k}, p_k\}$
goes into another probabilistic mixture $\{g(P_{\psi_k}), p_k\}$.
Therefore the two final density matrices after the action of $g$ on
two different probabilistic mixtures $\{P_{\psi_i}, p_i\}$ and
$\{P_{\phi_j}, q_j\}$ are
\be
\rho_A^\prime \{P_{\psi_i}, p_i\} = \sum \limits_i p_i
g(P_{\psi_i}) \nonumber
\\
\rho_A^\prime \{P_{\phi_j}, q_j\} = \sum \limits_j q_j
g(P_{\phi_j}) \label{rhoaprime} \ee which {\em a priori} can be
different. Let us recall that according to our assumptions the
results of all measurements in $A$ at a given time are determined
by the reduced density matrix $\rho_A^\prime$. This means that as a
consequence of the no-signaling principle the density matrix
$\rho_A^\prime$ at any later time has to be the same for all
probabilistic mixtures corresponding to a given initial density
matrix $\rho_A$. That is, it has to be {\em a function of $\rho_A$
only}.

We can therefore write
\be
\rho_A^\prime = g(\rho_A)=g(\sum \limits_i p_i P_{\psi_i}). \label{rhoaprime2} \ee
Eqs. (\ref{rhoaprime}) and (\ref{rhoaprime2}) together imply the
linearity of $g$:
\be
g(\sum \limits_i p_i P_{\psi_i})=\sum \limits_i p_i g(P_{\psi_i}).
\label{gofrho} \ee

Positivity of $g$ is necessary in order to ensure that $g(\rho_A)$
is again a valid density matrix, i.e. to ensure the positivity of
all probabilities calculated from it.

As we have made no specific assumptions about the system $A$ apart
from the fact that it can be entangled with some other spacelike
separated system, this means that the dynamics of our theory has to
be described by linear maps on density matrices in general.

Let us now argue that the {\em linearity} and {\em positivity}
already imply {\em complete positivity} in the present framework.
To see this, consider again two arbitrary subsystems $A$ and $B$
which may again be in an entangled state $|\psi\rangle_{AB}$. It is
conceivable that system $A$ is changed locally (i.e. the system
evolves, is measured etc.), which is described by some linear and
positive operation $g_A$, while {\em nothing} happens in $B$. This
formally corresponds to the operation $g_A\otimes\openone_B$ on the
whole system.

The joint operation $g_A\otimes\openone_B$ should take the density
matrix of the composite system $\rho_{AB}$ into another valid (i.e.
positive) density matrix, whatever the dimension of the system $B$.
But this is exactly the definition of {\em complete positivity} for
the map $g_A$ \cite{preskill}. If $g_A$ is positive but not CP,
then by definition there is always some entangled state $\rho_{AB}$
for which $g_A\otimes\openone_B$ applied to $\rho_{AB}$ is no
longer a positive density matrix and thus leads to unphysical
results such as negative probabilities.

In this way the existence of entangled states and the requirements
of positivity and linearity force us to admit only completely
positive dynamics. As mentioned already in the introduction, this
is equivalent to saying that under the given assumptions quantum
dynamics is essentially the {\em only} option since any CP map can
be realized by a quantum mechanical process, while on the other
hand, any quantum-mechanical process corresponds to a CP map
\cite{stochastic}.

There are three crucial ingredients in our argument: the existence
of entanglement, the trace rule, and the no-signaling condition.
Specifically, the trace rule leads to the preparation at a distance
of probabilistic mixtures and thus, as it were, to the right-hand
side of Eq. (\ref{gofrho}). On the other hand, the no-signaling
condition tells us that the dynamics is allowed to depend only on
the reduced density matrix, which leads to the left-hand side of
Eq. (\ref{gofrho}). Strictly speaking, in the derivation of
complete positivity, we have also used the assumption that the
identity operation on a subsystem is a permitted dynamical
evolution.

Nonlinear modifications of quantum mechanics
\cite{czachor,czachorkuna} have to give up at least one of these
assumptions. For instance, if the dynamics is allowed to depend on
the reduced density matrix $\rho_A$ in a nonlinear way, then it is
clear that $\rho_A$ cannot correspond to a probabilistic mixture of
pure states, cf. \cite{czachor}. But $\rho_A$ will correspond to
such a mixture whenever the observer in $B$ chooses to make
appropriate measurements, as long as we believe in the trace rule,
according to our above argument. This seems to imply that, at least
for separated systems, the trace rule has to be modified in such a
nonlinear theory.

Another example would be a theory where some entangled states are
{\em a priori} excluded. In this case some non-CP maps might be
permissible. An extreme example would be a theory without
entanglement. Such a theory would of course be in conflict with
experiments. An example for a linear, positive, but non-CP map
consistent with the no-signaling condition is the transposition of
the density matrix of the whole Universe (physically corresponding
to a time reversal). However in this case the identity operation on
a subsystem is not an allowed dynamics.

For completeness, let us finally show that any mixture
corresponding to a given density matrix can be prepared at a
distance from any entangled state with the correct reduced density
matrix \cite{gisinhpa,hughston}. Let us denote the system under
consideration by $A$ and the remote system by $B$.  Let us denote
the eigenvector representation of $\rho_A$ by $\sum_{k=1}^r
\lambda_k|v_k\rangle\langle v_k|$. Since the joint state
$|\psi\rangle_{A B}$ needs to have the correct reduced density
matrix, it must have a Schmidt decomposition
\be
|\psi\rangle_{A B}=\sum_{k=1}^r \sqrt{\lambda_k}\,|v_k\rangle
|g_k\rangle, \label{given} \ee where the $|g_k\rangle$ are
orthonormal states of system $B$. We want to show that {\it any}
decomposition of $\rho_A$ as a mixture of pure states can be
prepared from this state by operations on system $B$ only. To this
end, consider an arbitrary decomposition $\rho_A= \sum_{i=1}^m
x_i|\psi_i\rangle\langle\psi_i|$, where in general $m>r$. Clearly
this decomposition could be obtained from a state
\be
|\phi\rangle_{A
B}=\sum_{i=1}^m\sqrt{x_i}\,|\psi_i\rangle|\alpha_i\rangle,
\label{desired} \ee with the $|\alpha_i\rangle$ being an orthonormal basis of
a $m$--dimensional Hilbert space $H_m$. It seems that we now
require a larger Hilbert space in location $B$ in order to
accommodate all the orthonormal $|\alpha_i\rangle$. But the state
$|\phi\rangle_{A B}$ must also have a Schmidt representation
\be
|\phi\rangle_{A B}=\sum_{k=1}^r\sqrt{\lambda_k}\,|v_k\rangle
|h_k\rangle, \ee with $|h_k\rangle$ being orthonormal states in
$B$. This implies that $|\phi\rangle_{A B}$ and $|\psi\rangle_{A
B}$ are connected by a unitary transformation on $B$ alone:
\be
 |\phi\rangle_{A B}=\openone_A\otimes U_B |\psi\rangle_{A B}.
\ee The dimension of the support of the reduced density matrix
$\rho_B$ is the same for both states, since it is given by the
dimension of the support of $\rho_A$.

Thus one can prepare $|\phi\rangle_{AB}$ from any state with the
correct reduced density matrix by extending the system $B$ locally
to $m$ dimensions (using an appropriate ancilla), and then perform
the required (Von Neumann) measurement in the basis of the
$|\alpha_i\rangle$. This will correspond to a generalized
measurement \cite{peres} on the original $r$-dimensional system. In
this way every possible decomposition of $\rho_A$ can be prepared
at a distance.

In conclusion, we have shown that the basic dynamical rule of
quantum physics can be derived from its static properties and the
condition of no superluminal communication. This result puts
significant constraints on non-linear modifications of quantum
physics. It is clearly difficult to modify just parts of the whole
structure. More universal departures from the formalism may still
be possible without violating the no-signaling condition. We would
like to mention related recent work by Mielnik \cite{mielnik}.

C.S. would like to thank \v{C}. Brukner, L. Hardy and L. Vaidman
for useful discussions. This work was supported by the Austrian
Science Foundation FWF, project number S6503, and the TMR, QIPC and
EQUIP (IST-1999-11053) programs of the European Union.


\begin{references}
\bibitem{nlobs} One can say that a non-linear time evolution leads to the existence of
additional observables, which are defined
by letting the system evolve for a certain period of time, and then
measuring some standard quantum observable.
\end{references}
\end{document}